\documentclass[journal,final,onecolumn,12pt,letterpaper]{IEEEtran}%
\pdfoutput=1
\usepackage{amsmath}
\usepackage{amsfonts}
\usepackage{cite}
\usepackage{amssymb}
\usepackage{graphicx}%
\providecommand{\U}[1]{\protect\rule{.1in}{.1in}}

\begin{document}

\title{Connectivity of Confined Dense Networks: Boundary Effects and Scaling Laws}
\author{Justin P. Coon,~Carl P. Dettmann, and Orestis Georgiou\thanks{J. P. Coon is
with Telecommunications Research Laboratory, Toshiba Research Europe Ltd., 32
Queen Square, Bristol, UK, BS1 4ND; tel: +44 (0)117 906 0700, fax: +44 (0)117
906 0701, email: justin@toshiba-trel.com. }}
\maketitle

\begin{abstract}
In this paper, we study the probability that a dense network confined within a
given geometry is fully connected. We employ a cluster expansion approach
often used in statistical physics to analyze the effects that the boundaries
of the geometry have on connectivity. To maximize practicality and
applicability, we adopt four important point-to-point link models based on
outage probability in our analysis: single-input single-output (SISO),
single-input multiple-output (SIMO), multiple-input single-output (MISO), and
multiple-input multiple-output (MIMO). Furthermore, we derive diversity and
power scaling laws that dictate how boundary effects can be mitigated (to
leading order) in confined dense networks for each of these models. Finally,
in order to demonstrate the versatility of our theory, we analyze boundary
effects for dense networks comprising MIMO point-to-point links confined
within a right prism, a polyhedron that accurately models many geometries that
can be found in practice. We provide numerical results for this example, which
verify our analytical results.

\end{abstract}


\section{Introduction}

Multihop relay networks have received a lot of attention recently due to their
ability to improve coverage and, thus, capacity in a geographical sense. Many
of these networks -- such as mesh networks, vehicular networks, wireless
sensor networks, and \emph{ad hoc} networks -- possess commonality insomuch as
the number and distribution of nodes in the network is often random. A
considerable amount of research on random networks has been conducted in the
past (see, e.g., \cite{Balister2008,Haenggi2009,Li2009}). From a
communications perspective, it is of paramount importance to understand the
connectivity properties of such networks. This understanding can lead to
improved protocols and network deployment methodologies in practice
\cite{Ravelomanana2004}.

In recent years, researchers have adopted and adapted a number of techniques
from the physics and mathematical communities to study connectivity in random
networks. Perhaps the most directly applicable theory that has been employed
is that of continuum percolation \cite{Meester1996}, owing to its long and
rich history of use in describing particle clustering in statistical physics
and fluid dynamics
\cite{Hill1955,Coniglio1977a,Coniglio1977,Chiew1983,Stell1996}. In particular,
percolation theory is concerned with the emergence of a single large connected
component (possibly in addition to other finite connected components) in a
large (typically unbounded) graph, and dictates the minimum node density --
which is known as the \emph{critical node density} -- required to obtain such
a component. The links to network connectivity are obvious. Percolation theory
has been applied in recent years to identify and analyze power management
techniques that can be used to ensure network connectivity
\cite{Glauche2003,Paschos2009}. It has also been used to demonstrate the
benefits that node cooperation gives to improving connectivity
\cite{Goeckel2009,Westphal2009,Liu2010,Capar2011}. Further applications of the
general theory of percolation (including bond and site percolation) can be
found in studies of network resilience \cite{Callaway2000}, hybrid networks
(i.e., random networks with a regular element) \cite{Dousse2002}, coverage and
connectivity in wireless sensor networks \cite{Ammari2008} (a direct
application of \cite{Chiew1983}), and the information theoretic capacity of
networks \cite{Franceschetti2007}.

While the benefits of using percolation theory to explore asymptotic
connectivity issues in random networks is clear, the theory does not directly
address the question: what is the probability that \emph{all} nodes in the
network are connected? This question falls under the heading of \emph{full
connectivity} rather than percolation. The answer to this question is, of
course, related to a number of parameters, such as the fading environment, the
path loss model, the node density, and the geometry in which the network
resides. Many researchers have studied full connectivity, typically in some
asymptotic regime. Two particular network models have been popularized for the
study of full connectivity: the \emph{extended network model} and the
\emph{dense network model}. The former relates to the case where the node
density is finite and the network area is large, whereas the latter specifies
a finite network area with a high node density (see, e.g., \cite{Mao2011} and
the references therein).

With regard to extended networks, in \cite{Goeckel2009,Capar2011}, the authors
assume a \emph{unit disk connection model}\footnote{This model specifies that
connection between two nodes is achieved if and only if the distance between
them is at most $r$, which is some fixed positive number.} and give conditions
on the path loss exponent for which full connectivity can be achieved in
one-dimensional and two-dimensional networks. For dense networks, a number of
scaling laws have been published. For example, in \cite{Gupta1998}, the
authors derive a power scaling law that ensures full connectivity is achieved
\emph{almost surely} as the number of nodes in the network tends to infinity.
In \cite{Xue2004}, scaling laws are given for the number of nearest neighbors
(i.e., connections per node) that are required to achieve full connectivity
asymptotically in the number of nodes. Related results are given in
\cite{Xue2006} for sectorized networks.

More practical connection models (as opposed to the unit disk model) have also
been considered in the literature
\cite{Orriss2003,Bettstetter2004,Bettstetter2005,Hekmat2006,Miorandi2008,Mao2011a}%
. Specifically, the authors of \cite{Bettstetter2004,Bettstetter2005}
considered a probabilistic connection model whereby log-normal shadowing was
incorporated into the path loss model. Both 1-connectivity and $k$%
-connectivity\footnote{A $k$-connected network is one that remains fully
connected when $k^{\prime}<k$ point-to-point connections are broken.} were
considered in those papers. In \cite{Miorandi2008}, the authors considered the
effects of fast fading (in particular, Rayleigh and Rician fading) on full
connectivity in dense networks, and derived expressions for the node isolation
probability for several cases of interest. Note that this probability
effectively defines a first order approximation of the full-connectivity
probability for dense networks, a point that we will elaborate upon later.
Finally, the authors of \cite{Mao2011a} have recently studied connectivity
with respect to the decay properties of a general \emph{pair-connectedness
probability} \emph{function}.

\begin{figure}
[ptb]
\begin{center}
\includegraphics[
natheight=7.708100in,
natwidth=29.375099in,
height=1.6086in,
width=6.0494in
]%
{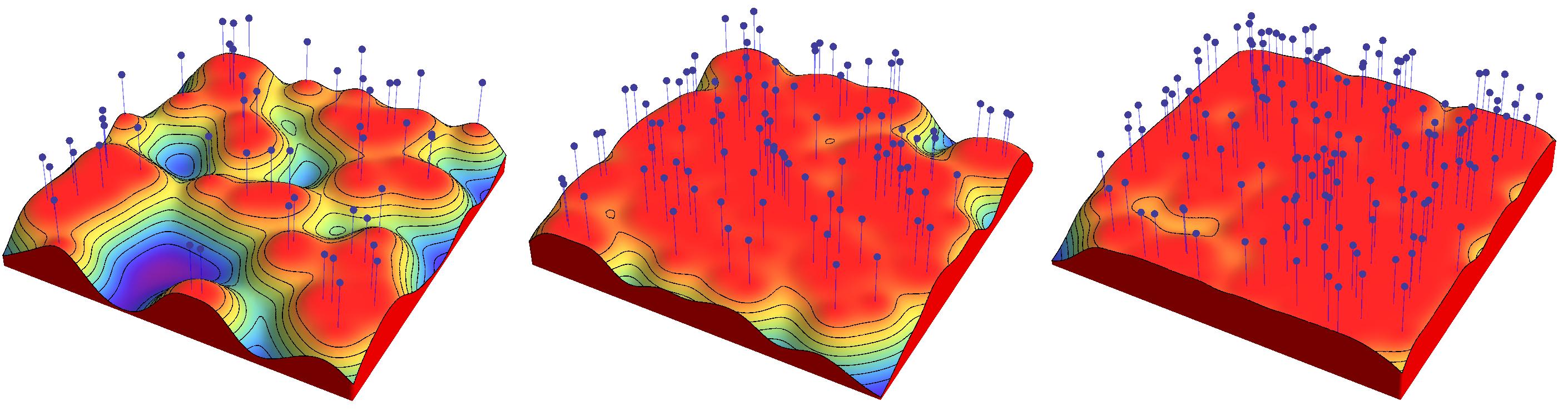}%
\caption{An illustration of the effects that the boundaries of the network
domain have on connectivity. In this example, the network domain is a square
of side length $L=10$ in arbitrary units. The node densities are 0.5, 1, and
1.5, reading left to right. The positions of the nodes for each depicted
realization are signified by pins, while the total probability of a new node
introduced to the network connecting to any other node is indicated by the
contours and shading (blue valleys indicate a low probability of connection
while red peaks indicate a high probability). Notice that at high densities,
low connection probabilities are concentrated near the corners.}%
\label{fig:illus}%
\end{center}
\end{figure}

The various contributions related to dense networks typically specify some
geometry in which the network resides. These geometries are commonly taken to
be squares or circles in two dimensions. One aspect that is common to
virtually all of this research is that boundary effects are neglected in the
interest of deriving a simple elegant result. \emph{In this paper, we show
that such effects should not, in general, be neglected since they tend to
dictate performance (in terms of the full-connectivity probability) in the
high density limit\footnote{See Fig. \ref{fig:illus} for a qualitative
illustration of this concept.}.} We provide a constructive explanation of this
assertion through the application of a novel cluster expansion model, first
developed in \cite{Coon2012,Coon2012a}, which admits an accurate first order
inhomogeneous approximation in the limit of large density. Such models arise
frequently in statistical physics to study the interaction between particles.
The application of this approach to the problem of network connectivity
carries many advantages, which together lead to the novel contributions of our
work. These contributions can be summarized as follows:

\begin{enumerate}
\item We utilize the inhomogeneous cluster expansion model to analyze the
effects that boundaries have on connectivity for four important point-to-point
link models: single-input single-output (SISO), single-input multiple-output
(SIMO), multiple-input single-output (MISO), and multiple-input
multiple-output (MIMO). To facilitate this analysis, we define the notion of
the \emph{mass of connectivity}, which is an average measure of pair
connectedness in a given volume taking boundaries into account.

\item We derive diversity and power scaling laws that dictate how boundary
effects can be mitigated (to leading order) in confined dense networks. These
laws are given for all four of the aforementioned point-to-point link models.

\item We corroborate the general theory detailed in \cite{Coon2012} by
analyzing boundary effects for dense networks comprising MIMO point-to-point
links confined within a right prism\footnote{A \emph{right prism} is a
polyhedron constructed by taking an $n$-sided polygon as its base, replicating
it and translating it in the \textquotedblleft vertical\textquotedblright%
\ direction, then connecting the corresponding sides. Thus, right prisms are
representative of, for example, many room configurations in a building.}. This
example is both instructive and insightful, illustrating the versatility and
advantages gained by employing the new cluster expansion theory in network
analysis problems.
\end{enumerate}%

The rest of the paper is organized as follows. We begin in Section
\ref{sec:pair_connectedness} by giving some preliminary details, which include
the point-to-point path loss model and pair-connectedness probabilities for
SISO, SIMO, MISO, and MIMO links. In Section \ref{sec:connectivity}, we give
details of the cluster expansion model and derive diversity and power scaling
laws for mitigating boundary effects to leading order. We then go one step
further by analyzing the connectivity of a network confined in a right prism.
Analytical and numerical results for a specific right prism are given in
Section \ref{sec:numerics}. Finally, we draw some conclusions in Section
\ref{sec:conc}.

\section{Preliminaries\label{sec:pair_connectedness}}

In this section, we provide preliminary details of the pair connectedness
model that we adopt. Specifically, we have that nodes $i$ and $j$ are directly
connected with probability $H(d(\mathbf{r}_{i},\mathbf{r}_{j}))$, which we
write as $H\left(  \mathbf{r}_{ij}\right)  $ or just $H_{ij}$, where the
distance function $d(\mathbf{r}_{i},\mathbf{r}_{j})$ is non-negative, zero
only when $\mathbf{r}_{i}=\mathbf{r}_{j}$, and symmetric. We define the pair
connectedness probability $H_{ij}$ as the complement of the information outage
probability between nodes $i$ and $j$. This well-understood metric is a
natural choice for pair connectedness since it provides fundamental insight
into network connectivity behavior, and can even be transformed directly into
a unit disk model through appropriate parameter definition if so desired;
however, it should also be noted that other pair connectedness models can
easily be chosen, such as a model based on the average bit-error rate of a
point-to-point link. The outage probability is generally parameterized by the
received signal-to-noise ratio (SNR), which in turn is a function of many
system parameters as well as the path loss. Consequently, we build our model
of pair connectedness starting from a fundamental definition of the path loss
model we employ.

\subsection{Path Loss Model and Outage Probability}

The received power of an electromagnetic wave decreases with distance like
$r^{-\eta}$ where $\eta$ is an environment-dependent decay parameter.
Typically, $\eta=2$ if propagation occurs in free space, with $\eta>2$ in
cellular/cluttered environments or through walls (see, \emph{e.g.},
\cite{Paulraj2003} and references therein). It follows that the SNR at the
receiver (assuming a fixed transmit power $P_{T}$ and a sufficiently narrow
bandwidth\footnote{Wideband channels sometimes exhibit different path loss
behavior, which varies as a function of frequency. Discussion of these
channels is beyond the scope of this paper; the interested reader is referred
to \cite{Kunisch2002,Cepeda2007} and references therein.}) also decays like
$r^{-\eta}$. Now, the outage probability for the link between nodes $i$ and
$j$ is defined as the probability that the $ij$ link cannot support a given
rate $R_{o}$ in bits per complex dimension, which can be written as%
\[
P_{ij}=\Pr\left(  \log_{2}\left(  1+\mathsf{SNR}\cdot X_{ij}\right)
<R_{o}\right)
\]
where $X_{ij}$ denotes the random variable signifying the power of the channel
between nodes $i$ and $j$. The pair connectedness probability is simply
$H_{ij}=1-P_{ij}$, which, after some manipulation, yields%
\begin{equation}
H_{ij}\left(  r\right)  =1-F_{X_{ij}}\left(  \beta r^{\eta}\right)
\label{eq:Hij}%
\end{equation}
where $F_{X_{ij}}$ is the cummulative distribution function of $X_{ij}$, and
$\beta$ is a constant -- which depends on the frequency of the transmission,
the power of the noise process at the receiver, and the transmit power -- that
defines the length scale. It is important to note that $\beta$ is inversely
proportional to the average received SNR. Finally, we point out that by
letting $\eta\rightarrow\infty$, we obtain the unit disk connection model:%
\begin{equation}
H_{ij}\left(  r\right)  =\left\{
\begin{array}
[c]{ll}%
1, & r<1\\
1-F_{X_{ij}}\left(  \beta\right)  , & r=1\\
0, & r>1
\end{array}
\right.  .
\end{equation}

\subsection{Point-to-Point Link Models}

We consider four general point-to-point link models: SISO, SIMO, MISO, and
MIMO links. For simplicity, we consider the case where individual channel
fading distributions follow a Rayleigh model, and all channels are
statistically independent. It follows that $X_{ij}$ has a standard exponential
distribution in the SISO case and a chi-squared distribution with $2m$ degrees
of freedom in the SIMO/MISO cases (where $m$ is the number of diversity
branches employed) \cite{Tse2005}. Thus, for SISO links, we have%
\begin{equation}
H_{ij}\left(  r\right)  =e^{-\beta r^{\eta}}\label{eq:HijSISO}%
\end{equation}
and for SIMO/MISO links, we have%
\begin{equation}
H_{ij}\left(  r\right)  =\frac{\Gamma\left(  m,\beta r^{\eta}\right)  }%
{\Gamma\left(  m\right)  }\label{eq:HijSIMO}%
\end{equation}
where $\Gamma\left(  a,x\right)  $ is the upper incomplete gamma function.
Equation (\ref{eq:HijSIMO}) reduces to the SISO case when $m=1$.

For MIMO links with $n_{t}$ transmit antennas and $n_{r}$ receive antennas, we
are mostly concerned with ensuring connectivity is achieved. Thus, we assume
beamforming is applied at the transmitter of each node while maximum ratio
combining (MRC) is employed at the receiver \cite{Kang2003}. For this
so-called MIMO MRC channel, the pair connectedness probability becomes%
\begin{equation}
H_{ij}\left(  r\right)  =1-\kappa_{m,n}\det\left(  \gamma\left(
n-m+i+j-1,\beta r^{\eta}\right)  \right)  _{ij}\label{eq:HijMIMO}%
\end{equation}
where $n=\max\left(  n_{t},n_{r}\right)  $, $m=\min\left(  n_{t},n_{r}\right)
$, $i,j=1,\ldots,m$, $\gamma\left(  a,x\right)  $ is the lower incomplete
gamma function, and%
\[
\kappa_{m,n}=\left(  \prod_{i=1}^{m}\Gamma\left(  n-i+1\right)  \Gamma\left(
m-i+1\right)  \right)  ^{-1}.
\]
In order to aid analysis, we will focus on the special case where $m=2$. This
restriction is justified on the basis of pragmatism: indeed, one would
envisage that low complexity may be a requirement of nodes operating in dense
networks. This is certainly the case in wireless sensor networks where sensors
are often powered by batteries \cite{Li2009}. In any case, we maintain some
level of generality by not restricting $n$. For $m=2$ and general $n$, we can
express $H_{ij}$ as%
\begin{equation}
H_{ij}\left(  r\right)  =1-nP\left(  n-1,\beta r^{\eta}\right)  P\left(
n+1,\beta r^{\eta}\right)  +\left(  n-1\right)  P\left(  n,\beta r^{\eta
}\right)  ^{2}\label{eq:HijMIMOm2}%
\end{equation}
where $P\left(  n,x\right)  $ is the regularized lower incomplete gamma
function. We will use these formulae for $H_{ij}$ in the next section to
analyze the probability that a network is connected at the boundary of the
confining geometry.

\section{Probability of Full Connectivity\label{sec:connectivity}}

In this section, we develop a novel theory of the probability of full network
connectivity in confined geometries through the use of a cluster expansion
technique. Cluster expansions are frequently used in statistical mechanics and
fluid dynamics to study the interaction between particles. In our application
of network connectivity, we first give an overview of the model, then provide
details of the first order expansion, which can be used to study connectivity
in the high density limit for general pair-connectedness models. For more
details on this model, the interested reader is referred to \cite{Coon2012a}.
We then utilize the pair-connectedness functions given in the previous section
to analyze the probability of full connectivity at the boundary of the
confining geometry to leading order, providing rules that dictate how
diversity and power can be scaled to mitigate boundary effects.

\subsection{First Order Cluster Expansion}

Consider $N$ randomly distributed nodes with locations $\mathbf{r}_{i}%
\in\mathcal{V}\subseteq\mathbb{R}^{d}$ for $i=1,2,\ldots,N$ according to a
uniform density $\rho=N/V$, where $V=\left\vert \mathcal{V}\right\vert $ and
$\left\vert \cdot\right\vert $ denotes the size of the set. Here, we use the
Lebesgue measure of the appropriate dimension $d$. We define the average of a
quantity as%
\[
\left\langle A\right\rangle =\frac{1}{V^{N}}\int_{\mathcal{V}^{N}}A\left(
\mathbf{r}_{1},\mathbf{r}_{2},\ldots,\mathbf{r}_{N}\right)  \mathrm{d}%
\mathbf{r}_{1}\mathrm{d}\mathbf{r}_{2}\cdots\mathrm{d}\mathbf{r}_{N}.
\]

We define some useful notation. Let $S=\{1,2,3,\ldots,N\}$. A graph $g=\left(
A,L\right)  $ consists of a set $A\subseteq S$ of nodes, together with a
collection $L\subseteq\left\{  (i,j)\in A:i<j\right\}  $ of direct links, that
is unordered distinct pairs of nodes. As a slight abuse of notation, we write
$(i,j)\in g$ to denote that $(i,j)$ is an element of the set of links $L$
associated with the graph $g$. We write $G^{A}$ for the set of graphs with
nodes in $A$, and $G_{j}^{A}$ for the set with nodes in $A$ and largest
connected component (cluster) of size $j$ with $1\leq j\leq\left\vert
A\right\vert $.

The probability of two nodes being connected or not leads to the trivial
identity%
\[
1=H_{ij}+(1-H_{ij}).
\]
Multiplying over all links with nodes in a set $A$ expresses the probability
of all possible combinations. This can be written as
\begin{equation}
1=\prod_{i,j\in A;\,i<j}\left(  H_{ij}+\left(  1-H_{ij}\right)  \right)
=\sum_{g\in G^{A}}\mathcal{H}_{g}\label{eq:expansion}%
\end{equation}
where
\begin{equation}
\mathcal{H}_{g}=\prod_{(i,j)\in g}H_{ij}\prod_{(i,j)\not \in g}\left(
1-H_{ij}\right)  .
\end{equation}
The sum in (\ref{eq:expansion}) contains $2^{\left\vert A\right\vert \left(
\left\vert A\right\vert -1\right)  /2}$ separate terms. Setting $A=S$, this
can be expressed as collections of terms determined by their largest cluster,
which yields
\begin{equation}
1=\underbrace{\sum_{g\in G_{N}^{S}}\mathcal{H}_{g}}_{P_{fc}\left(
\mathbf{r}_{1},\ldots,\mathbf{r}_{N}\right)  }+\sum_{g\in G_{N-1}^{S}%
}\mathcal{H}_{g}+\cdots+\underbrace{\sum_{g\in G_{1}^{S}}\mathcal{H}_{g}%
}_{\prod_{i<j}\left(  1-H_{ij}\right)  }\label{eq:expansion2}%
\end{equation}

For a given configuration of node positions $\mathbf{r}_{i}\in\mathcal{V}$,
assuming that the nodes are pairwise connected with independent probabilities
$H_{ij}$, the first term in equation~(\ref{eq:expansion2}) is the probability
of being fully connected $P_{fc}(\mathbf{r}_{1},\ldots\mathbf{r}_{N})$. The
average of this quantity over all possible configurations $P_{fc}=\left\langle
P_{fc}(\mathbf{r}_{1},\ldots\mathbf{r}_{N})\right\rangle $ is the overall
probability of obtaining a fully connected network and is our desired quantity
of interest. Hence, rearranging equation~(\ref{eq:expansion2}) allows us to
obtain expressions for $P_{fc}$ in a consistent way while keeping track of
correction terms.

Note that in the high density limit of $\rho\rightarrow\infty$, the right hand
side of (\ref{eq:expansion2}) is dominated by the first term, which yields
$P_{fc}\approx1$, and hence the network is fully connected with probability
one as expected. The approximation symbol is used here to indicate that first
and higher order corrections are being ignored.

The first order approximation is obtained by expanding the second term on the
right-hand side of (\ref{eq:expansion2}) explicitly, taking into account all
possible ways of getting an $N-1$ cluster. Thus the average probability that
an $N$-node network confined in $\mathcal{V}$ is fully connected (to first
order) is
\begin{align}
P_{fc} &  \approx1-\left\langle \sum_{g\in G_{N-1}^{S}}\mathcal{H}%
_{g}\right\rangle \nonumber\\
&  =1-\left\langle \left(  \sum_{\wp=1}^{N}\prod_{j\neq\wp}\left(  1-H_{j\wp
}\right)  \right)  \underbrace{\left(  \sum_{g\in G_{N-1}^{S\setminus\left\{
\wp\right\}  }}\mathcal{H}_{g}\right)  }_{\approx1}\right\rangle \nonumber\\
&  =1-N\left\langle \prod_{j=1}^{N-1}\left(  1-H_{jN}\right)  \right\rangle
\nonumber\\
&  =1-\frac{N}{V^{N}}\int_{\mathcal{V}^{N}}\prod_{j=1}^{N-1}\left(  1-H\left(
\mathbf{r}_{jN}\right)  \right)  \mathrm{d}\mathbf{r}_{1}\cdots\mathrm{d}%
\mathbf{r}_{N}\label{eq:Pcaux}%
\end{align}
where the simplification in the third equality is valid since all nodes are
identical, meaning the sum over $\wp$ can be factored out.

At this point, we deviate from common practice and assume that the network is
\emph{not} translationally symmetric, \emph{i.e.}, the system is
inhomongeneous. This ensures that we will include boundary effects in our
analysis. Indeed, it turns out that it is not only important to include such
effects, but that they \emph{dictate} \emph{performance} in the limit of high
density. This is demonstrated to a large degree through the first order
cluster expansion. In particular, we can progress from (\ref{eq:Pcaux}) to
obtain%
\begin{align}
P_{fc} &  \approx1-\frac{N}{V}\int_{\mathcal{V}}\left(  1-\frac{1}{V}%
\int_{\mathcal{V}}H\left(  \mathbf{r}_{12}\right)  \mathrm{d}\mathbf{r}%
_{1}\right)  ^{N-1}\mathrm{d}\mathbf{r}_{2}\nonumber\\
&  =1-\rho\int_{\mathcal{V}}e^{-\rho\int_{\mathcal{V}}H\left(  \mathbf{r}%
_{12}\right)  \mathrm{d}\mathbf{r}_{1}}\left(  1+O\left(  N^{-1}\right)
\right)  \mathrm{d}\mathbf{r}_{2}\label{eq:Pfc}%
\end{align}
in the limit of large $N$. In fact, this equation was recently given in
\cite{Mao2011a} (eq. (8)), where $V$ was scaled exponentially with $\rho$,
which effectively implies boundary effects are ignored. For our approximation,
however, we only require that $V\gg\rho$, or equivalently $V\gg\sqrt{N}$ (cf.
\cite{Coon2012a} for more details on scaling limits of our theory). This is a
key difference from previous approaches reported in the literature.

Equation (\ref{eq:Pfc}) suggests that in the high density limit, the
probability of having a single $N-1$ connected cluster (\emph{i.e.}, having an
isolated node) is dominated by nodes that are situated in \textquotedblleft
hard to connect\textquotedblright\ regions of the available domain
$\mathcal{V}$. This is because the outer integral in (\ref{eq:Pfc}) is
dominated where the integral in the exponential is small, which occurs at
corners, edges, and faces. For nodes located near these geometric effects, the
volume in range of the nodes is small.

\subsection{Pair Connectedness and Scaling Laws\label{sec:scaling_laws}}

Taking a closer look at the integral in the exponent in (\ref{eq:Pfc}), we see
that it effectively defines the \emph{mass of connectivity} for a given
pair-connectedness function at $\mathbf{r}_{2}$, and we label it accordingly:%
\begin{equation}
M_{H}\left(  \mathbf{r}_{2}\right)  =\int_{\mathcal{V}}H\left(  \mathbf{r}%
_{12}\right)  \mathrm{d}\mathbf{r}_{1}.
\end{equation}
Indeed, $M_{H}$ can be studied in detail for different pair-connectedness
functions, and power and diversity scaling laws for bounded network
connectivity can be gleaned from such an analysis. In the discussion that
follows, it will be understood that $M_{H}$ is a function of $\mathbf{r}_{2}$,
and thus we will refrain from explicitly writing out this relationship for the
sake of brevity.

We now incorporate the pair-connectedness functions given in Section
\ref{sec:pair_connectedness} into the model. A common feature of these
connection functions is that they decay exponentially with $r$. This allows us
to separate the $\mathrm{d}\mathbf{r}_{1}$ integral in (\ref{eq:Pfc}) and
extract its leading order behavior by supposing that $\mathbf{r}_{2}$ is
located somewhere on the boundary $\partial\mathcal{V}$ of the network domain
$\mathcal{V}\subset\mathbb{R}^{d}$ \cite{Coon2012a}. Thus, for a general pair
connectedness probability, we have to leading order\footnote{We emphasize that
equality is only to leading order here.}%
\begin{equation}
M_{H}=\underbrace{\left(  \int_{0}^{\infty}r^{d-1}H_{ij}\left(  r\right)
\mathrm{d}r\right)  }_{M_{H}^{\prime}}\left(  \int\mathrm{d}\Omega\right)
\label{eq:boundary_integral}%
\end{equation}
where $\Omega=2\pi^{d/2}/\Gamma\left(  d/2\right)  $ is the full solid angle
in $d$ dimensions, $\omega=\int\mathrm{d}\Omega\in\left(  0,\Omega\right)  $
is the solid angle as seen from $\mathbf{r}_{2}$, and we call $M_{H}^{\prime}$
the \emph{homogeneous mass of connectivity} since it characterizes the mass of
connectivity in a homogeneous system. Note that the domain of integration in
the $\mathrm{d}r$ integral is not truncated from above because the integrand
is exponentially decaying and, thus, does not contribute significantly to the
integral for large arguments (\emph{i.e.}, for nodes located far from the node
at $\mathbf{r}_{2}$). This approximation clearly demonstrates the influence
that boundaries have on connectivity. Specifically, by writing%
\begin{equation}
M_{H}=\delta\Omega M_{H}^{\prime}%
\end{equation}
where $\delta=\omega/\Omega\in\left(  0,1\right)  $, we see that connectivity
is dominated from regions that are hard to connect and is controlled by the
solid angle available to them. However, we cannot ignore the contribution of
$M_{H}^{\prime}$, which will be seen to be useful in determining ways to
mitigate a small solid angle contribution.

We now investigate the leading order behavior for the four link models
detailed in Section \ref{sec:pair_connectedness}, placing particular emphasis
on how diversity and power control can be used to mitigate boundary effects.
Note that in order for us to obtain a more detailed view of boundary effects,
we must expand (\ref{eq:boundary_integral}) beyond leading order; a discussion
to this effect is given in Section \ref{sec:prisms}.

\subsubsection{SISO, SIMO, and MISO Link Models}

For SISO, SIMO, and MISO pair connectedness, we have%
\begin{equation}
M_{H}^{\prime}=\int_{0}^{\infty}r^{d-1}\frac{\Gamma\left(  m,\beta r^{\eta
}\right)  }{\Gamma\left(  m\right)  }\mathrm{d}r
\end{equation}
The $\mathrm{d}r$ integral can be calculated directly by using the standard
integral definition of the incomplete gamma function%
\[
\Gamma\left(  m,x\right)  =\int_{x}^{\infty}t^{m-1}e^{-t}\mathrm{d}t
\]
and exchanging the order of integration, which eventually yields%
\begin{equation}
M_{H}^{\prime}=\frac{\Gamma\left(  m+\frac{d}{\eta}\right)  }{\beta^{\frac
{d}{\eta}}d\Gamma\left(  m\right)  }.\label{eq:leading_order_SIMO}%
\end{equation}
For uncluttered ($\eta=2$) $d$-dimensional networks with point-to-point SISO
links, $m=1$ and we have the simple result%
\[
M_{H}=\delta\left(  \frac{\pi}{\beta}\right)  ^{\frac{d}{2}}.
\]

\begin{figure}
[ptb]
\begin{center}
\includegraphics[
natheight=7.135600in,
natwidth=10.239400in,
height=3.525in,
width=5.047in
]%
{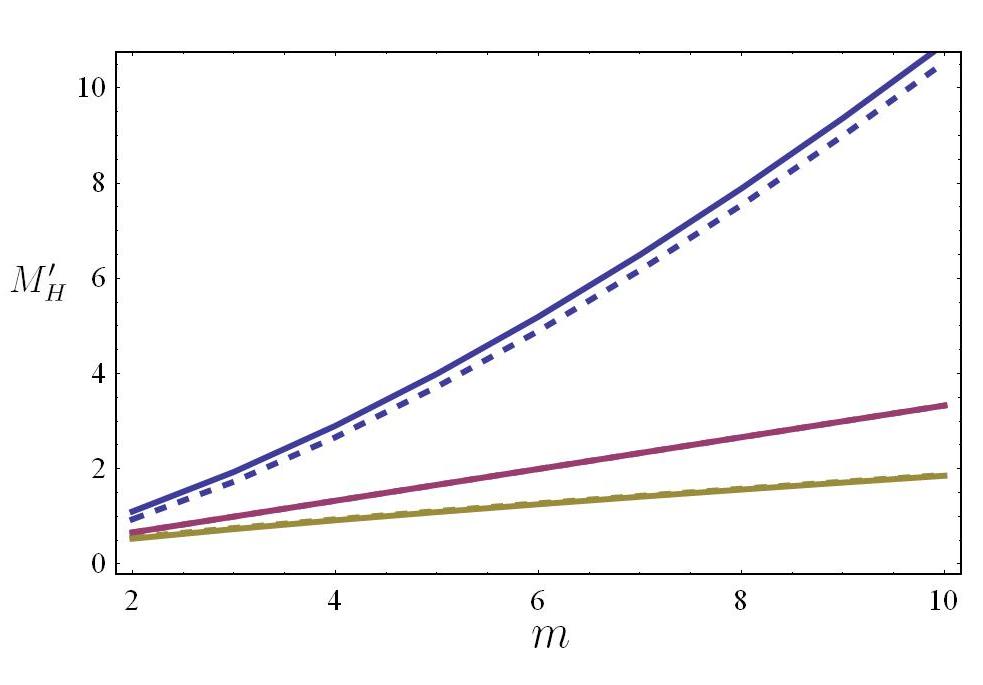}%
\caption{An illustration of the $M_{H}^{\prime}$ scaling law for SIMO/MISO
systems. The exact expression given by (\ref{eq:leading_order_SIMO}) is
represented by solid lines for $d=3$ and $\eta=2,3,4$ moving from top to
bottom. The dashed curves correspond to the leading order behavior.}%
\label{fig:simo}%
\end{center}
\end{figure}

Intuitively, one might expect that boundary effects can be mitigated somewhat
through the use of diversity (dictated by $m$ in this case). Expanding
(\ref{eq:leading_order_SIMO}) to leading order as $m$ grows large yields%
\begin{equation}
M_{H}^{\prime}=\frac{m^{\frac{d}{\eta}}}{\beta^{\frac{d}{\eta}}d}\left(
1+O\left(  m^{-1}\right)  \right)  .
\end{equation}
Thus, we see that although an increase in $m$ results in better connectivity,
the rate of increase is highly dependent upon the dimension of the space in
which the network resides and the path loss exponent. Indeed, by scaling $m$,
it is possible to obtain a progressive improvement in performance in
uncluttered environments ($\eta\approx2$), whereas cluttered environments
($\eta>3$) will typically yield diminishing returns. To illustrate this
scaling law, we have plotted the exact expression for $M_{H}^{\prime}$ along
with the leading order term in Fig. \ref{fig:simo}. Not only is the leading
order behavior apparent in this example, but we see that the leading order, in
fact, provides an excellent approximation.%

As a final note, recall that $\beta$ is inversely proportional to the average
received SNR at a given node. Consequently, it can be seen from
(\ref{eq:leading_order_SIMO}) that an increase in the transmit power for each
node has the same scaling effect as an increase in the diversity order $m$.
However, it should be noted that this analysis is only valid when such an
increase is less than exponential relative to the system size, \emph{i.e.},
scaling the transmit power must not counteract the exponential decay in pair connectedness.

The fact that boundary effects can be somewhat mitigated through the
appropriate use of diversity and/or power scaling is a powerful conclusion
with significant practical implications, particularly since these scaling laws
have identical order. For example, some networks may be energy/power
constrained (\emph{e.g.}, wireless sensor networks and cognitive networks)
whereas others may be constrained by form size (\emph{e.g.}, some vehicular
networks). The scaling laws presented here provide engineering insight into
how to design networks for a plethora of scenarios.

\subsubsection{MIMO Link Models}

We can perform the same analysis for MIMO systems. In this case, we restrict
our attention to systems where $m=2$, for which $H_{ij}$ can be rewritten from
(\ref{eq:HijMIMOm2}) into the form%
\begin{multline}
H_{ij}\left(  r\right)  =\frac{n\Gamma\left(  n-1,\beta r^{\eta}\right)
-2\Gamma\left(  n,\beta r^{\eta}\right)  +\left(  n-1\right)  ^{-1}%
\Gamma\left(  n+1,\beta r^{\eta}\right)  }{\Gamma\left(  n-1\right)  }\\
+\frac{\Gamma\left(  n,\beta r^{\eta}\right)  ^{2}-\Gamma\left(  n-1,\beta
r^{\eta}\right)  \Gamma\left(  n+1,\beta r^{\eta}\right)  }{\Gamma\left(
n\right)  \Gamma\left(  n-1\right)  }.
\end{multline}
Using this form of $H_{ij}$, we can evaluate $M_{H}^{\prime}$ by integrating
term by term and using recurrence relations of hypergeometric functions to
yield%
\begin{multline}
M_{H}^{\prime}=\frac{\left(  1-\frac{d}{\eta}\right)  \Gamma\left(
n-1+\frac{d}{\eta}\right)  }{\beta^{\frac{d}{\eta}}d\Gamma\left(  n-1\right)
}+\frac{\eta\Gamma\left(  2n+\frac{d}{\eta}\right)  }{\beta^{\frac{d}{\eta}%
}d\Gamma\left(  n\right)  ^{2}}\Bigg(\frac{1}{n}F\left(  n-1,2n+\frac{d}{\eta
};n+1;-1\right) \\
-\frac{n-1}{\left(  n+\frac{d}{\eta}\right)  \left(  n-1+\frac{d}{\eta
}\right)  }F\left(  n-1+\frac{d}{\eta},2n+\frac{d}{\eta};n+1+\frac{d}{\eta
};-1\right)  \Bigg)\label{eq:leading_order_MIMO2}%
\end{multline}
where $F\left(  a,b;c;z\right)  $ denotes the hypergeometric function. The
terms involving hypergeometric functions result from integrating the quadratic
terms in the expression for $H_{ij}$ given above. Again, the effect of the
solid angle defining the volume in range is apparent. Although the expression
given above is quite cumbersome and does not necessarily lead to much
intuitive insight, we can easily evaluate the approximation for specific
values of $n$. For example, we have for $n=2$%
\begin{equation}
M_{H}^{\prime}=\frac{\left(  \left(  \frac{d}{\eta}\right)  ^{2}+\frac{d}%
{\eta}+2-2^{-\frac{d}{\eta}}\right)  \Gamma\left(  \frac{d}{\eta}\right)
}{\beta^{\frac{d}{\eta}}\eta}.
\end{equation}

\begin{figure}
[ptb]
\begin{center}
\includegraphics[
natheight=7.135600in,
natwidth=10.166700in,
height=3.55in,
width=5.047in
]%
{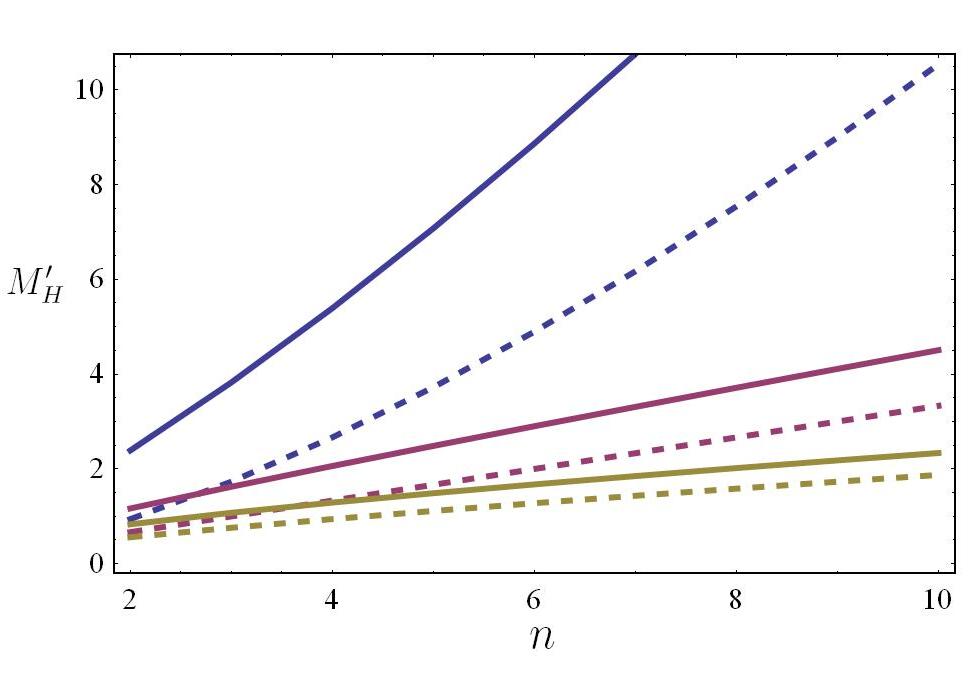}%
\caption{An illustration of the $M_{H}^{\prime}$ scaling law for MIMO systems.
The exact expression given by (\ref{eq:leading_order_MIMO2}) is represented by
solid lines for $d=3 $ and $\eta=2,3,4$ moving from top to bottom. The dashed
curves correspond to the leading order behavior.}%
\label{fig:mimo}%
\end{center}
\end{figure}

On the other hand, we can obtain a scaling law in $n$ by noting that for large
$n$, the contributions of the $P$-functions in (\ref{eq:HijMIMOm2}) are not
significant for $\beta r^{\eta}<n$. Moreover, for $\beta r^{\eta}>n$,
$H_{ij}\left(  r\right)  $ decays exponentially. Thus, we can approximate
$H_{ij}\left(  r\right)  $ by a step function with a transition at $r=\left(
n/\beta\right)  ^{\frac{1}{\eta}}$. We present an argument for making this
approximation in Appendix \ref{app:stepfunctionapprox}. It follows that we can write%

\begin{equation}
M_{H}^{\prime}=\int_{0}^{\left(  \frac{n}{\beta}\right)  ^{\frac{1}{\eta}}%
}r^{d-1}\mathrm{d}r+\epsilon\left(  n\right)  =\frac{1}{d}\left(  \frac
{n}{\beta}\right)  ^{\frac{d}{\eta}}+\epsilon\left(  n\right)
\end{equation}
where the error term is given by%
\begin{equation}
\epsilon\left(  n\right)  =\underbrace{\int_{0}^{\left(  \frac{n}{\beta
}\right)  ^{\frac{1}{\eta}}}r^{d-1}\left(  H_{ij}\left(  r\right)  -1\right)
\mathrm{d}r}_{\epsilon_{-}\left(  n\right)  }+\underbrace{\int_{\left(
\frac{n}{\beta}\right)  ^{\frac{1}{\eta}}}^{\infty}r^{d-1}H_{ij}\left(
r\right)  \mathrm{d}r}_{\epsilon_{+}\left(  n\right)  }.
\end{equation}
Note that $\epsilon_{-}\left(  n\right)  $ corresponds to the negative
contribution and $\epsilon_{+}\left(  n\right)  $ is the positive
contribution. It remains to determine the order (in $n$) of the error term. In
fact, it can be shown (by way of accurate approximation if not rigorously)
that%
\begin{equation}
\epsilon\left(  n\right)  =O\left(  n^{\frac{d}{\eta}-\frac{1}{2}}\right)  .
\end{equation}
A derivation of this result is given in Appendix \ref{app:errorterm}. Thus, we
can write%
\begin{equation}
M_{H}^{\prime}=\frac{n^{\frac{d}{\eta}}}{\beta^{\frac{d}{\eta}}d}\left(
1+O\left(  n^{-\frac{1}{2}}\right)  \right)
\end{equation}
which is completely analogous to the SIMO/MISO case. This scaling law is
illustrated in Fig. \ref{fig:mimo}, where we have plotted the exact expression
for $M_{H}^{\prime}$ given by (\ref{eq:leading_order_MIMO2}) along with the
leading order term. The leading order growth in $n$ is apparent in this
example, but in contrast to the SIMO/MISO case, it is not a good
approximation. This results from the fact that the first correction term is
$O\left(  n^{-1/2}\right)  $ rather than $O\left(  n^{-1}\right)  $ as is the
case in SIMO/MISO systems. Finally, it is easy to see (by observing
(\ref{eq:leading_order_MIMO2})) that the same transmit power scaling law
holds, \emph{i.e.}, $M_{H}^{\prime}$ scales like $\beta^{-\frac{d}{\eta}}$.%

\section{General Formula and Applications to Right Prisms\label{sec:prisms}}

In \cite{Coon2012a}, we developed a general formula for studying the
full-connectivity probability in the limit of high density for any network
domain. This formula takes the form%
\begin{equation}
P_{fc}\approx1-\sum_{i=0}^{d}\sum_{j_{i}}\rho^{1-i}G_{j_{i}}V_{j_{i}}%
e^{-\rho\omega_{j_{i}}\int_{0}^{\infty}r^{d-1}H\left(  r\right)  \mathrm{d}%
r}\label{eq:general_formula}%
\end{equation}
where, again, $d$ is the dimension, $G_{j_{i}}$ is a geometrical factor for
each object $j_{i}$ of codimension $i$, and $V_{j_{i}}$ is the corresponding
$d-i$ dimensional volume of the object with solid angle $\omega_{j_{i}}$. In
the spirit of this formula, we now derive an expression for $P_{fc}$ by
evaluating (\ref{eq:Pfc}) when the network in question is located in a right
prism and the point-to-point links are modeled as uncluttered (\emph{i.e.},
$\eta=2$) $2\times2$ MIMO channels. The choice of MIMO point-to-point links
for this analysis was taken to demonstrate that our methodology can be applied
to reasonably complicated scenarios. Indeed, SISO, SIMO and MISO links can
also be analyzed.

The pair connectedness probability for $2\times2$ MIMO channels with $\eta=2$
simplifies to%
\begin{equation}
H\left(  r\right)  =e^{-\beta r^{2}}\left(  \beta^{2}r^{4}+2-e^{-\beta r^{2}%
}\right)
\end{equation}
where $r$ is the distance between two nodes. To derive $P_{fc}$ in the form of
(\ref{eq:general_formula}), we evaluate the integral (cf. (\ref{eq:Pfc}))%
\begin{equation}
\int_{\mathcal{V}}e^{-\rho\int_{\mathcal{V}}H\left(  \mathbf{r}_{12}\right)
\mathrm{d}\mathbf{r}_{1}}\mathrm{d}\mathbf{r}_{2}=\int_{\mathcal{V}}e^{-\rho
M_{H}\left(  \mathbf{r}_{2}\right)  }\mathrm{d}\mathbf{r}_{2}%
\label{eq:integral_for_prism}%
\end{equation}
for each local feature of the geometry\footnote{Here, we have explicitly
indicated that the mass of connectivity $M_{H}$ is dependent upon
$\mathbf{r}_{2}$ since we will be concerned with first order correction
terms.}. This illustrative example will further show how geometric effects
influence connectivity. Moreover, the choice of a right prism as the confining
geometry demonstrates the power and versatility of our theory, particularly
since many practical geometries can be well approximated by such a polyhedron.

At this point, it is beneficial and instructive to give an outline of the
approach we use to analyze networks confined in right prisms. Our general
method is to begin by considering features with the lowest dimension,
\emph{i.e.}, corners in this case. We then move to edges, then faces, and
finally the bulk of the prism. At each step, we ignore objects of lower
dimension since we have accounted for their contribution to connectivity in
previous steps. It will be observed that this is a particularly powerful
approach when analyzing the effects that the bulk and faces have since the
surface (volume) of a right prism is locally equivalent to that of a sphere of
the appropriate surface area (volume). We now give details of our analysis,
and follow this discussion with a specific example.

\subsection{Corners}

For a right prism, each corner is the product of three intersecting planes
that can be oriented such that at least one edge connected to its vertex is
normal to an adjoining face. This suggests that we should use cylindrical
coordinates to perform calculations for this case, where we locate the origin
at the vertex and the $z$-axis is oriented along the edge that connects the
two identical polygons.

The distance between two points in cylindrical coordinates is given by%
\[
d\left(  \mathbf{r}_{1},\mathbf{r}_{2}\right)  =\sqrt{r_{1}^{2}+r_{2}%
^{2}-2r_{1}r_{2}\cos\left(  \theta_{1}-\theta_{2}\right)  +\left(  z_{1}%
-z_{2}\right)  ^{2}}%
\]
where $\left(  r_{i},\theta_{i},z_{i}\right)  $ are the coordinates of a node
located at $\mathbf{r}_{i}$. Thus, in order to evaluate the inner integral of
(\ref{eq:integral_for_prism}) near the corner, we let $\mathbf{r}_{2}$ be
located near the corner and expand $H\left(  \mathbf{r}_{12}\right)  $ near
$r_{2}=0$ and $z_{2}=0$, which, to first order, yields%
\begin{multline}
H\left(  \mathbf{r}_{12}\right)  =e^{-2\beta\left(  r_{1}^{2}+z_{1}%
^{2}\right)  }\left(  e^{\beta\left(  r_{1}^{2}+z_{1}^{2}\right)  }\left(
2+\beta^{2}\left(  r_{1}^{2}+z_{1}^{2}\right)  ^{2}\right)  -1\right) \\
+2\beta e^{-2\beta\left(  r_{1}^{2}+z_{1}^{2}\right)  }\left(  e^{\beta\left(
r_{1}^{2}+z_{1}^{2}\right)  }\left(  2+\beta\left(  r_{1}^{2}+z_{1}%
^{2}\right)  \left(  \beta\left(  r_{1}^{2}+z_{1}^{2}\right)  -2\right)
\right)  -2\right) \\
\times\left(  z_{1}z_{2}+r_{1}r_{2}\cos\left(  \theta_{1}-\theta_{2}\right)
\right)  .\label{eq:H_expanded_0}%
\end{multline}
We can now perform the inner integrals in (\ref{eq:integral_for_prism}) as
follows:%
\begin{align}
M_{H}\left(  \mathbf{r}_{2}\right)   & =\int_{0}^{\infty}\int_{0}^{\vartheta
}\int_{0}^{\infty}r_{1}H\left(  \mathbf{r}_{12}\right)  \mathrm{d}%
r_{1}\mathrm{d}\theta_{1}\mathrm{d}z_{1}\nonumber\\
& =\frac{1}{8\beta}\left(  14z_{2}\vartheta+\frac{23-\sqrt{2}}{2}\sqrt
{\frac{\pi}{\beta}}\vartheta+7\pi r_{2}\left(  \sin\theta_{2}-\sin\left(
\theta_{2}-\vartheta\right)  \right)  \right) \label{eq:inner_integral_corner}%
\end{align}
where $\vartheta$ is the angle of the corner with $0<\vartheta<\pi$. Note that
semi-infinite integration is allowed here due to the fact that $H$ is
exponentially decreasing but the system size is large\footnote{In particular,
if $L$ denotes the typical length of the geometry, then we require
$\sqrt{\beta}L\gg1$.}. Thus, contributions to the integral at distant
boundaries are negligible. Using (\ref{eq:inner_integral_corner}), we are now
in a position to calculate the outer integrals of (\ref{eq:integral_for_prism}%
), which yields%
\begin{align}
\int_{\mathcal{V}}e^{-\rho M_{H}\left(  \mathbf{r}_{2}\right)  }%
\mathrm{d}\mathbf{r}_{2}  & =\iiint r_{2}e^{-\rho\int_{\mathcal{V}}H\left(
\mathbf{r}_{12}\right)  \mathrm{d}\mathbf{r}_{1}}\mathrm{d}r_{2}%
\mathrm{d}z_{2}\mathrm{d}\theta_{2}\nonumber\\
& =\frac{256\beta^{3}\csc\vartheta}{343\pi^{2}\rho^{3}\vartheta}%
e^{-\frac{\left(  23-\sqrt{2}\right)  \sqrt{\pi}\rho\vartheta}{16\beta^{3/2}}%
}.\label{eq:corner}%
\end{align}
The regions of integration are the same here as for the inner integrals;
however, note that the order of integration changes.

All that remains is to enumerate the $2n$ corners for a prism constructed from
an $n$-sided polygon. Each corner is defined by the angle $\vartheta$. For
example, a cuboid, which is a right prism formed by replicating and
translating a square, has eight corners, all of which have angle
$\vartheta=\pi/2$.

\subsection{Edges}

Now we consider geometric features of dimension one: edges. Let $L$ be the
length of the edge in question. The calculations for this case are also
facilitated by the use of cylindrical coordinates, but where the origin is
located at the center of the edge. Thus, the corners are located at $\pm L/2$
and the angle of the corner is $\vartheta$. Since we wish to ignore effects
from corners, faces, and the bulk, we expand $H$ about $r_{2}=0$ and $z_{2}=0
$, which gives (\ref{eq:H_expanded_0}). Calculating the inner integrals in
(\ref{eq:integral_for_prism}) to first order yields an expression that has
$\exp\left(  -\beta L^{2}/4\right)  $ and $\operatorname{erf}\left(
L\sqrt{\beta/2}\right)  $ terms. Again, assuming that $\sqrt{\beta}L\gg1$, we
can make the approximations $\exp\left(  -\beta L^{2}/4\right)  \approx0$ and
$\operatorname{erf}\left(  L\sqrt{\beta/2}\right)  \approx1$. It follows that
$M_{H}\left(  \mathbf{r}_{2}\right)  $ can be evaluated to yield%
\begin{equation}
M_{H}\left(  \mathbf{r}_{2}\right)  =\frac{1}{4\beta}\left(  \frac{23-\sqrt
{2}}{2}\sqrt{\frac{\pi}{\beta}}\vartheta+7\pi r_{2}\left(  \sin\theta_{2}%
-\sin\left(  \theta_{2}-\vartheta\right)  \right)  \right)
\end{equation}
where the integrals are performed over $r_{1}\in\left(  0,\infty\right)  $,
$\theta_{1}\in\left(  0,\vartheta\right)  $, and $z_{1}\in\left(
-L/2,L/2\right)  $. The outer integrals in (\ref{eq:integral_for_prism}) can
now be performed to yield%
\begin{equation}
\int_{-\frac{L}{2}}^{\frac{L}{2}}\int_{0}^{\vartheta}\int_{0}^{\infty}%
r_{2}e^{-\rho M_{H}\left(  \mathbf{r}_{2}\right)  }\mathrm{d}r_{2}%
\mathrm{d}\theta_{2}\mathrm{d}z_{2}=\frac{16L\beta^{2}\csc\vartheta}{49\pi
^{2}\rho^{2}}e^{-\frac{\left(  23-\sqrt{2}\right)  \sqrt{\pi}\rho\vartheta
}{8\beta^{3/2}}}.\label{eq:edges}%
\end{equation}
Again, all that remains is to enumerate the $3n$ edges.

\subsection{Faces}

For the contribution of the faces to the full-connectivity probability, we
employ a \emph{local equivalence} argument that allows us to greatly simplify
the analysis. Specifically, we have covered the corner and edge calculations
above, and thus we ignore contributions from these features when considering
faces. Thus, one can imagine deforming a prism of surface area $S$ into a
sphere of the same surface area, the radius $R$ of which is defined by the
relation $S=4\pi R^{2}$. For a general right prism, the surface area is given
by%
\[
S=2B+ph
\]
where $B$ is the area of the base (\emph{i.e.}, the $n$-sided polygon), $p$ is
the base perimeter, and $h$ is the height. If the base is a regular $n$-sided
polygon with side length $s$, we have%
\[
S=\frac{n}{2}s^{2}\cot\frac{\pi}{n}+nsh.
\]
Thus, this argument allows us to treat any convex right prism that we
wish\footnote{Convexity is required since we only consider line-of-sight
connections between nodes, although small-scale scattering is accounted for
through the chosen pair-connectedness models.}.

Using spherical coordinates along with the fact that the distance between
nodes at $\mathbf{r}_{1}$ and $\mathbf{r}_{2}$ is given by%
\[
d\left(  \mathbf{r}_{1},\mathbf{r}_{2}\right)  =\sqrt{r_{1}^{2}+r_{2}%
^{2}-2r_{1}r_{2}\cos\theta}%
\]
where $r_{i}=\left\vert \mathbf{r}_{i}\right\vert $ and $\theta\in\left[
0,\pi\right]  $ is the angle between the nodes, we expand $H$ near the surface
of the sphere (\emph{i.e.}, $r_{2}=R$) and perform the inner integrals in
(\ref{eq:integral_for_prism}) to obtain%
\begin{align}
M_{H}\left(  \mathbf{r}_{2}\right)   & =2\pi\int_{0}^{R}\int_{0}^{\pi}%
r_{1}^{2}\sin\theta\,H\left(  \mathbf{r}_{12}\right)  \mathrm{d}%
\theta\mathrm{d}r_{1}\nonumber\\
& =\frac{\pi}{4\beta}\left(  \frac{23-\sqrt{2}}{2}\sqrt{\frac{\pi}{\beta}%
}+14\left(  R-r_{2}\right)  \right)
\end{align}
where the factor of $2\pi$ follows from integration over the azimuthal angle.
To arrive at this result, it was assumed that $\sqrt{\beta}R\gg1$, which
allows us to make similar approximations to the error functions of the form
$\operatorname{erf}\left(  c\sqrt{\beta}R\right)  $ and exponentials of the
form $\exp\left(  -c\beta R^{2}\right)  $ for some constant $c>0$ as was done
for edges. Furthermore, we have made the additional approximations
$c/R\approx0$ and $c/R^{2}\approx0$. The outer integrals can now be performed
to yield (to dominant term in $R$ and $\rho$)%
\begin{equation}
2\pi\int_{0}^{R}\int_{0}^{\pi}r_{2}^{2}\sin\theta~e^{-\rho M_{H}\left(
\mathbf{r}_{2}\right)  }\mathrm{d}\theta\mathrm{d}r_{2}=\frac{8\beta R^{2}%
}{7\rho}e^{-\frac{\left(  23-\sqrt{2}\right)  \pi^{3/2}\rho}{8\beta^{3/2}}}.
\end{equation}
Generalizing this result to any right prism, \emph{i.e.}, substituting $S=4\pi
R^{2}$, gives%
\begin{equation}
\int_{\mathcal{V}}e^{-\rho M_{H}\left(  \mathbf{r}_{2}\right)  }%
\mathrm{d}\mathbf{r}_{2}=\frac{2\beta S}{7\pi\rho}e^{-\frac{\left(
23-\sqrt{2}\right)  \pi^{3/2}\rho}{8\beta^{3/2}}}.\label{eq:surfaces}%
\end{equation}
The faces do not need to be enumerated in this case since we have accounted
for all faces through the local equivalence argument.

Finally, it is worth mentioning that face contributions can also be calculated
in a lengthy manner by using cartesian coordinates. This works well for
rectangular sides; however, one must be more careful when considering more
general $n$-sided polygons. In any case, it is straightforward to show for
certain simple cases that the proposed approach to the calculation yields
identical results to the more complex cartesian approach.

\subsection{Bulk}

For the bulk contribution, we apply the same local equivalence argument that
was used for the face contributions, but where the expansion in $H$ is
performed at $r_{2}=0$. In other words, we consider a sphere of radius $R$
determined by the relation $V=\frac{4}{3}\pi R^{3}$, where%
\[
V=Bh
\]
is the volume of the right prism containing the network. For a prism
constructed from a base that is a regular $n$-sided polygon with side length
$s$, we have%
\[
V=\frac{n}{4}hs^{2}\cot\frac{\pi}{n}.
\]

Expanding $H$ about the origin to first order, we can perform the inner
integration to obtain%
\begin{align}
M_{H}\left(  \mathbf{r}_{2}\right)   & =2\pi\int_{0}^{\infty}\int_{0}^{\pi
}r_{1}^{2}\sin\theta\,H\left(  \mathbf{r}_{12}\right)  \mathrm{d}%
\theta\mathrm{d}r_{1}\nonumber\\
& =\frac{\left(  23-\sqrt{2}\right)  }{4}\left(  \frac{\pi}{\beta}\right)
^{\frac{3}{2}}.
\end{align}
Finally, the outer integrals can be evaluated to yield%
\begin{equation}
2\pi\int_{0}^{R}\int_{0}^{\pi}r_{2}^{2}\sin\theta~e^{-\rho M_{H}\left(
\mathbf{r}_{2}\right)  }\mathrm{d}\theta\mathrm{d}r_{2}=Ve^{-\frac{\left(
23-\sqrt{2}\right)  \pi^{3/2}\rho}{4\beta^{3/2}}}.\label{eq:volume}%
\end{equation}

\subsection{Example: the \textquotedblleft House\textquotedblright\ Prism
\label{sec:numerics}}

\begin{figure}
[ptb]
\begin{center}
\includegraphics[
natheight=10.645800in,
natwidth=9.885700in,
height=4.3517in,
width=4.0421in
]%
{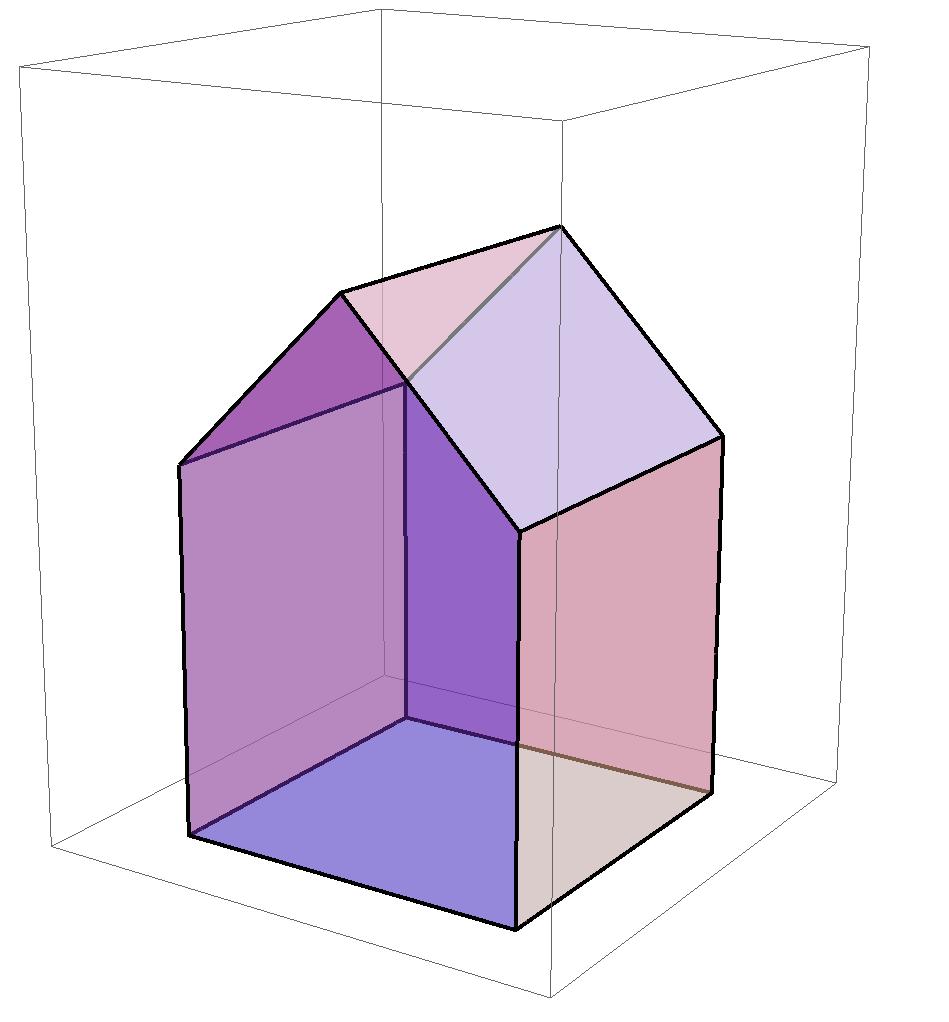}%
\caption{The \textquotedblleft house\textquotedblright\ prism considered for
numerical results. The base is a square of side $L$, the apex is a right
angle, and the total height is $3L/2$.}%
\label{fig:house}%
\end{center}
\end{figure}

We now apply the calculations detailed above to a particular scenario.
Specifically, we consider a network comprised of $2\times2$ MIMO links with
$\eta=2$ operating in a right prism that resembles a typical \textquotedblleft
house\textquotedblright, as illustrated in Fig. \ref{fig:house}. The base of
the house is a square of side $L$. The total height of the prism is $3L/2$,
and the apex is right angled. This example somewhat illustrates the
versatility of our theory, particularly since most previously published
results have considered very simple geometries, such as squares and circles in
two dimensions, or cubes and spheres in three dimensions. It should be noted
that this is an arbitrary example, and in fact many more complicated
geometries can be analyzed with the proposed methodology.%

Analysis of the \textquotedblleft house\textquotedblright\ prism is
straightforward using the calculations given in the preceding sections. We
first identify the different corner angles and count the multiplicities of
each. There are ten corners in this prism, six of which have angle
$\vartheta=\pi/2$ and four of which have angle $\vartheta=3\pi/4$. Thus, the
following two contributions to the general formula for $P_{fc}$ arise from the
corners (cf. (\ref{eq:corner})):%
\begin{equation}
C_{1}=6\frac{512\beta^{3}}{343\pi^{3}\rho^{3}}e^{-\frac{\left(  23-\sqrt
{2}\right)  \rho}{32}\left(  \frac{\pi}{\beta}\right)  ^{3/2}}%
\end{equation}
and%
\begin{equation}
C_{2}=4\frac{1024\sqrt{2}\beta^{3}}{1029\pi^{3}\rho^{3}}e^{-\frac{\left(
23-\sqrt{2}\right)  3\rho}{64}\left(  \frac{\pi}{\beta}\right)  ^{3/2}}.
\end{equation}

Next, we perform a similar step for the edges, of which there are fifteen.
Thirteen edges are right angled: nine of these have length $L$ while the
remaining four have length $L/\sqrt{2}$. The other two edges have angle
$\vartheta=3\pi/4$ and length $L$. Thus, we can write the following two edge
contributions to the high density expression for $P_{fc}$ (cf. (\ref{eq:edges}%
)):%
\begin{equation}
E_{1}=L\left(  9+2\sqrt{2}\right)  \frac{16\beta^{2}}{49\pi^{2}\rho^{2}%
}e^{-\frac{\left(  23-\sqrt{2}\right)  \rho}{16}\left(  \frac{\pi}{\beta
}\right)  ^{3/2}}%
\end{equation}
and%
\begin{equation}
E_{2}=2L\frac{16\sqrt{2}\beta^{2}}{49\pi^{2}\rho^{2}}e^{-\frac{\left(
23-\sqrt{2}\right)  3\rho}{32}\left(  \frac{\pi}{\beta}\right)  ^{3/2}}.
\end{equation}

For the faces of the prism, we calculate the total surface area to be%
\begin{equation}
S=\frac{11+2\sqrt{2}}{2}L^{2}.
\end{equation}
Thus, using the local equivalence argument presented above, we can substitute
this area into (\ref{eq:surfaces}) to obtain the contribution of all of the
faces to $P_{fc}$. We denote this contribution as $F$. Similarly, we can take
the volume of the prism to be%
\begin{equation}
V=\frac{5}{4}L^{3}.
\end{equation}
Substituting into (\ref{eq:volume}) yields the bulk contribution to $P_{fc}$,
which we denote by $U$.

Finally, the general formula is obtained through the summation of all
contributions, which yields%
\begin{equation}
P_{fc}\approx1-\rho\left(  C_{1}+C_{2}+E_{1}+E_{2}+F+U\right)  .
\end{equation}
It can easily be seen that this formula has the same structure as
(\ref{eq:general_formula}). In fact, the various contributions to the general
formula are outlined in Table \ref{tab:1}.%

\begin{table}[tbp] \centering
\caption{Contributions of the various geometrical features of the ``house''
prism to the general formula for $P_{fc}$ given by
(\ref{eq:general_formula}).  The angle $\vartheta = \pi / 2$ for type-1
corner/edge contributions, and $\vartheta = 3\pi / 4$ for type-2
contributions.}\label{tab:1}%
\begin{tabular}
[c]{|c||c|c|l|l|}\hline
General Formula Parameter & Corners & Edges & Faces & Bulk\\\hline\hline
Volume ($V_{j_{i}}$) & $1$ & $L,\,\frac{L}{\sqrt{2}}$ &
\multicolumn{1}{|c|}{$S$} & \multicolumn{1}{|c|}{$V$}\\
Solid Angle ($\omega_{j_{i}}$) & $\vartheta$ & $2\vartheta$ &
\multicolumn{1}{|c|}{$2\pi$} & \multicolumn{1}{|c|}{$4\pi$}\\
Geometrical Factor ($G_{j_{i}}$) & $\frac{256\beta^{3}\csc\vartheta}%
{343\pi^{3}\vartheta}$ & $\frac{16\beta^{2}\csc\vartheta}{49\pi^{2}}$ &
\multicolumn{1}{|c|}{$\frac{2\beta}{7\pi}$} & \multicolumn{1}{|c|}{$1$%
}\\\hline
\end{tabular}%
\end{table}%

\begin{figure}
[ptb]
\begin{center}
\includegraphics[
natheight=5.677500in,
natwidth=8.604000in,
height=3.2456in,
width=5.047in
]%
{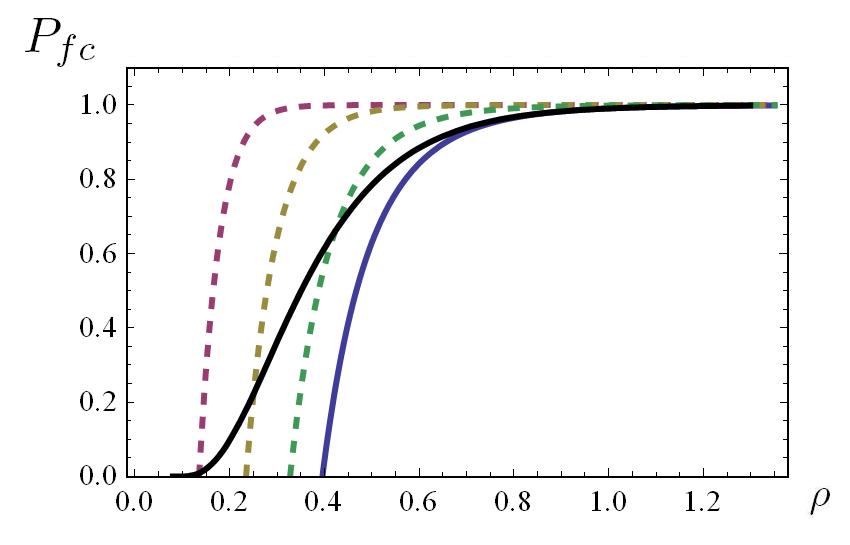}%
\caption{Analytical and numerical results for the connectivity of a network in
a typical \textquotedblleft house\textquotedblright. The two solid curves
(reading left to right) are the numerical and full analytical results. The
left-most dashed curve (red) is the general formula with only a bulk term
(\emph{i.e.}, face, edge, and corner terms are ignored). The next dashed curve
(yellow) includes bulk and face contributions. The right-most dashed curve
(green) includes bulk, face, and edge contributions, but not corners.}%
\label{fig:Pfc}%
\end{center}
\end{figure}

\begin{figure}
[ptb]
\begin{center}
\includegraphics[
natheight=5.677500in,
natwidth=8.374800in,
height=3.397in,
width=5.0462in
]%
{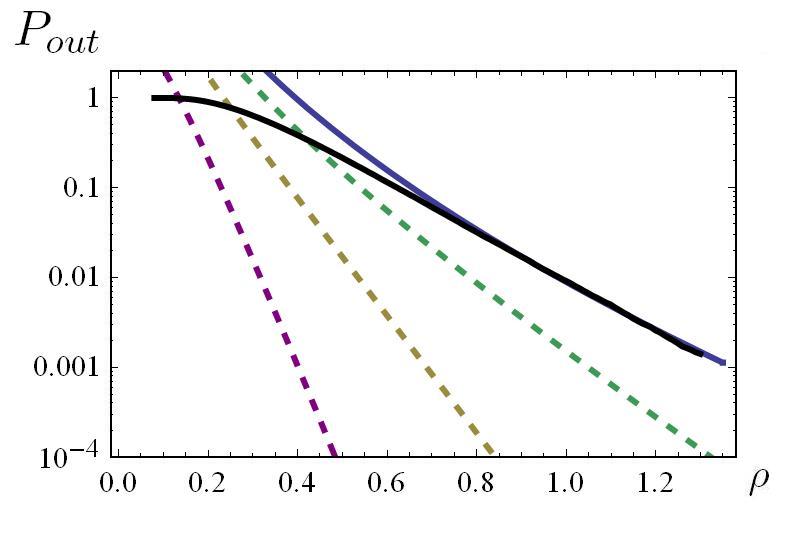}%
\caption{Analytical and numerical results for the network outage probability
in a typical \textquotedblleft house\textquotedblright. The labelling of the
curves is identical to Fig. \ref{fig:Pfc}.}%
\label{fig:Pout}%
\end{center}
\end{figure}

Letting $\beta=1$ and $L=7$ for simplicity, we have plotted the general
formula for this example along with numerical results obtained through
simulations in Fig. \ref{fig:Pfc}. Furthermore, to illustrate the accuracy of
our results at high density, we have plotted the probability of network outage
(\emph{i.e.}, $P_{out}=1-P_{fc}$) in Fig. \ref{fig:Pout}. In these figures,
the two solid curves (reading left to right in both figures) are the numerical
and full analytical results (including contributions from all geometrical
boundary features). Note that good agreement is achieved at high densities as
expected. The left-most dashed curve (red) is the general formula with only a
bulk term (\emph{i.e.}, all boundaries are ignored). This curve representes
\textquotedblleft conventional wisdom\textquotedblright, where boundary
effects are neglected, and serves as a benchmark for our results. This
benchmark is overly optimistic and should be treated with caution. This is
particularly clear in Fig. \ref{fig:Pout}. The next dashed curve (yellow)
includes bulk and face contributions. We see that by including even
two-dimensional boundary effects, the model is improved significantly. The
right-most dashed curve (green) includes bulk, face, and edge contributions,
but not corners.%

%

Although these results show that our theory is in excellent agreement with
simulations for high densities, they also illustrate the limitations of the
theory for low to mid-range densities. However, one notices that, for example,
boundaries are not as significant at low densities. Moreover, for mid-range
densities, one may only need to consider boundary effects due to faces and/or
edges, but not corners, in order to obtain accurate connectivity predictions.
This observation can be explained in, perhaps, two ways. Firstly, from a
mathematical point of view, our analysis relied heavily on expansions, both
near boundaries and at high density. To maintain simplicity and tractability,
these expansions were typically only to first order. This may be a limitation
when considering low to mid-range densities. Secondly, one may intuitively
expect that away from the region of high density there is a low probability of
a node being located \emph{very} close to a boundary, which of course is where
our analysis is valid (again, expansions were taken near boundaries). In any
case, our results actually suggest that the theory can be modified somewhat to
be valid in regions of low and mid-range densities. This is certainly of
interest, and is currently ongoing work.

\section{Conclusions\label{sec:conc}}

In this paper, we analyzed the probability of full connectivity for dense
networks confined within certain geometries. In contrast to much of the
results reported in the literature, we showed that boundary effects should not
be ignored, and in fact dictate performance in the high density regime. We
built upon our cluster expansion model of full network connectivity, first
reported in \cite{Coon2012}, by adopting four important point-to-point link
models for pair connectedness: SISO, SIMO, MISO, and MIMO. This approach
allowed us to study the effects that the boundaries of the confining geometry
have on practical network connectivity, and led to the derivation of diversity
and power scaling laws that dictate how these effects can be mitigated.
Finally, we provided a practical example by analyzing the full-connectivity
probability of a network comprising MIMO point-to-point links confined within
a right prism. This example demonstrated the versatility of our theory in
network analysis problems, and we hope that this theory will inspire many
researchers in their own problems related to network design and performance analysis.

\appendices

\section{Argument for Approximating $H_{ij}$ by a Step
Function\label{app:stepfunctionapprox}}

Let $H$ be given by%
\[
H\left(  x\right)  =1-nP\left(  n-1,x\right)  P\left(  n+1,x\right)  +\left(
n-1\right)  P\left(  n,x\right)  ^{2}.
\]
First, consider the case where $x\ll n$. Employing the asymptotic equivalence
\cite[8.11.5]{NIST2010}%
\[
P\left(  n,x\right)  \sim\frac{e^{n-x}}{\sqrt{2\pi n}}\left(  \frac{x}%
{n}\right)  ^{n}%
\]
we can write%
\[
H\left(  x\right)  \sim1-\frac{e^{2\left(  n-x\right)  }}{2\pi n}\left(
\frac{x}{n}\right)  ^{2n}%
\]
and thus $H\left(  x\right)  \rightarrow1$ as $n\rightarrow\infty$.

Now, consider the case where $x$ is close to but less than $n$. Let $x=\lambda
n<n$ for some $\lambda\approx1$. Using the asymptotic relations \cite[5.11.3,
8.11.6]{NIST2010}, we can write%
\[
P\left(  n,x\right)  =\frac{\gamma\left(  n,x\right)  }{\Gamma\left(
n\right)  }\sim\frac{\left(  n-x\right)  ^{-1}x^{n}e^{-x}}{\sqrt{2\pi
}n^{n-\frac{1}{2}}e^{-n}}=\frac{\left(  \lambda e^{1-\lambda}\right)  ^{n}%
}{\sqrt{2\pi n}\left(  1-\lambda\right)  }.
\]
But $0\leq\lambda e^{1-\lambda}<1$ for $0\leq\lambda<1$, and thus $P\left(
n,\lambda n\right)  \rightarrow0$ for fixed $\lambda<1$. It follows that
$H\left(  x\right)  \approx1$ for $x<n$ and $n$ large.

Finally, consider the case where $x$ is close to but greater than $n$. Let
$x=\lambda n>n$ for some $\lambda\approx1$. Now we employ the asymptotic
relations \cite[5.11.3, 8.11.7]{NIST2010}, which give%
\[
P\left(  n,x\right)  =1-\frac{\Gamma\left(  n,x\right)  }{\Gamma\left(
n\right)  }\sim1-\frac{\left(  \lambda e^{1-\lambda}\right)  ^{n}}{\sqrt{2\pi
n}\left(  \lambda-1\right)  }.
\]
But $0\leq\lambda e^{1-\lambda}<1$ for $\lambda>1$, and thus $P\left(
n,\lambda n\right)  \rightarrow1$ for fixed $\lambda>1$. It follows that
$H\left(  x\right)  \rightarrow0$ for $x>n$ as $n\rightarrow\infty$. Thus, it
is logical to make the stated step function approximation with a transition at
$x=n$.

\section{Order of the Error Term\label{app:errorterm}}

To determine the order of $\epsilon\left(  n\right)  $, we will make use of
the expansions \cite[5.11.3]{NIST2010}%
\begin{equation}
\Gamma\left(  n\right)  \sim\sqrt{2\pi}n^{n-\frac{1}{2}}e^{-n}\left(
1+O\left(  n^{-1}\right)  \right) \label{eq:exp1}%
\end{equation}
and \cite[8.11.12]{NIST2010}%
\begin{equation}
\Gamma\left(  n,n\right)  \sim\sqrt{\frac{\pi}{2}}n^{n-\frac{1}{2}}%
e^{-n}\left(  1+O\left(  n^{-\frac{1}{2}}\right)  \right)  .\label{eq:exp2}%
\end{equation}
Moreover, we have the following identity (which holds for any $a$, but where
the logic behind the proof is given for $a>0$):%
\begin{align}
\frac{\Gamma\left(  n+a,n\right)  }{\Gamma\left(  n\right)  } &  =\frac
{\Gamma\left(  n+a,n\right)  }{\Gamma\left(  n+a\right)  }\frac{\Gamma\left(
n+a\right)  }{\Gamma\left(  n\right)  }\nonumber\\
&  =\left(  \frac{\Gamma\left(  n+a,n+a\right)  }{\Gamma\left(  n+a\right)
}+\frac{1}{\Gamma\left(  n+a\right)  }\int_{n}^{n+a}t^{n+a-1}e^{-t}%
\mathrm{d}t\right)  \frac{\Gamma\left(  n+a\right)  }{\Gamma\left(  n\right)
}\nonumber\\
&  =\frac{1}{2}n^{a}\left(  1+O\left(  n^{-\frac{1}{2}}\right)  \right)
.\label{eq:exp3}%
\end{align}
Similarly, we have%
\begin{equation}
\frac{\Gamma\left(  2n+a,2n\right)  }{\Gamma\left(  n\right)  ^{2}}%
=\frac{2^{2\left(  n-1\right)  +a}}{\sqrt{\pi}}n^{a+\frac{1}{2}}\left(
1+O\left(  n^{-\frac{1}{2}}\right)  \right)  .\label{eq:exp4}%
\end{equation}

We begin by determining the order of%
\[
\epsilon_{+}\left(  n\right)  =\int_{\left(  \frac{n}{\beta}\right)
^{\frac{1}{\eta}}}^{\infty}r^{d-1}H\left(  r\right)  \mathrm{d}r=\frac{1}%
{\eta\beta^{\frac{d}{\eta}}}\int_{n}^{\infty}x^{\frac{d}{\eta}-1}\tilde
{H}\left(  x\right)  \mathrm{d}x
\]
where $\tilde{H}\left(  x\right)  =H\left(  \left(  x/\beta\right)  ^{1/\eta
}\right)  $. We can rewrite $\tilde{H}\left(  x\right)  $ to yield the
following bound on the pair connectedness function:%
\begin{align*}
\tilde{H}\left(  x\right)   &  =2Q\left(  n,x\right)  -Q\left(  n,x\right)
^{2}+\frac{x^{2n-1}e^{-2x}}{\Gamma\left(  n\right)  ^{2}}+\left(  1-Q\left(
n,x\right)  \right)  \frac{\left(  x-n\right)  x^{n-1}e^{-x}}{\Gamma\left(
n\right)  }\\
&  \leq\underbrace{2Q\left(  n,x\right)  +\frac{x^{2n-1}e^{-2x}}{\Gamma\left(
n\right)  ^{2}}+\frac{\left(  x-n\right)  x^{n-1}e^{-x}}{\Gamma\left(
n\right)  }}_{\tilde{H}_{b}\left(  x\right)  },\quad x\geq n
\end{align*}
where $Q\left(  n,x\right)  =1-P\left(  n,x\right)  =\Gamma\left(  n,x\right)
/\Gamma\left(  n\right)  $. Now we can write%
\begin{align*}
\epsilon_{+}\left(  n\right)   & \leq\frac{1}{\eta\beta^{\frac{d}{\eta}}}%
\int_{n}^{\infty}x^{\frac{d}{\eta}-1}\tilde{H}_{b}\left(  x\right)
\mathrm{d}x\\
& =\frac{1}{\eta\beta^{\frac{d}{\eta}}\Gamma\left(  n\right)  ^{2}}\left(
2^{1-2n-\frac{d}{\eta}}\Gamma\left(  2n-1+\frac{d}{\eta},2n\right)  \right. \\
& \qquad\left.  -\Gamma\left(  n\right)  \left(  2\frac{\eta}{d}n^{\frac
{d}{\eta}}\Gamma\left(  n,n\right)  +n\Gamma\left(  n-1+\frac{d}{\eta
},n\right)  -\left(  1+2\frac{\eta}{d}\right)  \Gamma\left(  n+\frac{d}{\eta
},n\right)  \right)  \right)  .
\end{align*}
Using the expansions given in (\ref{eq:exp1})-(\ref{eq:exp4}), it can be shown
that this bound behaves like $O\left(  n^{\frac{d}{\eta}-\frac{1}{2}}\right)
$.

Considering the negative error term, we have%
\[
\left\vert \epsilon_{-}\left(  n\right)  \right\vert =\frac{1}{\eta
\beta^{\frac{d}{\eta}}}\int_{0}^{n}x^{\frac{d}{\eta}-1}\tilde{H}_{c}\left(
x\right)  \mathrm{d}x
\]
where $\tilde{H}_{c}\left(  x\right)  =1-\tilde{H}\left(  x\right)  $, which
behaves like $o\left(  x^{2n-\delta}\right)  $ for small $\delta>0$ as
$x\rightarrow0$, and increases sharply near $x=n$ for large $n$ (see Appendix
\ref{app:stepfunctionapprox}). Thus, we propose a semi-rigorous bound on
$\tilde{H}_{c}\left(  x\right)  $ in the form of an exponential with a slope
at $x=n$ matched to that of $\tilde{H}_{c}\left(  x\right)  $, \emph{i.e.},%
\[
\tilde{H}_{c}\left(  x\right)  \leq a_{n}e^{-b_{n}\left(  n-x\right)  }%
\]
where $a_{n}$ and $b_{n}$ are constants that may depend on $n$. Note that
$\tilde{H}_{c}\left(  x\right)  $ increases with $x$ and, thus, has a maximum
in the interval $x\in\left[  0,n\right]  $ at $x=n$. Furthermore, $\tilde
{H}_{c}\left(  n\right)  $ decreases with $n$, so $\tilde{H}_{c}\left(
x\right)  $ is upper bounded by%
\[
\tilde{H}_{c}\left(  n=2\right)  =\frac{2\left(  \cosh2-3\right)  }{e^{2}}.
\]
Furthermore, the slope of $\tilde{H}_{c}\left(  x\right)  $ at $x=n$ is given
by%
\[
\tilde{H}_{c}^{\prime}\left(  n\right)  =\frac{n^{n}e^{-2n}\left(
2n^{n}+e^{n}\gamma\left(  n+1,n\right)  \right)  }{\Gamma\left(  n+1\right)
^{2}}.
\]
Matching the slope of the bound at $x=n$ and setting the bound equal to
$\tilde{H}_{c}\left(  n=2\right)  $ at the same point leads to the definitions
$a_{n}=\tilde{H}_{c}\left(  n=2\right)  $ and $b_{n}=a^{-1}\tilde{H}%
_{c}^{\prime}\left(  n\right)  $.

Making the substitution $t=n-x$ allows us to write%
\[
\left\vert \epsilon_{-}\left(  n\right)  \right\vert \leq\frac{a_{n}%
n^{\frac{d}{\eta}-1}}{\eta\beta^{\frac{d}{\eta}}}\int_{0}^{n}\left(
1-\frac{t}{n}\right)  ^{\frac{d}{\eta}-1}e^{-b_{n}t}\mathrm{d}t.
\]
By the mean value theorem, we have%
\begin{align*}
\left\vert \epsilon_{-}\left(  n\right)  \right\vert  & \leq\frac{a_{n}\xi
_{n}^{\frac{d}{\eta}-1}n^{\frac{d}{\eta}-1}}{\eta\beta^{\frac{d}{\eta}}}%
\int_{0}^{n}e^{-b_{n}t}\mathrm{d}t\\
& =\frac{a_{n}\xi_{n}^{\frac{d}{\eta}-1}n^{\frac{d}{\eta}-1}}{b_{n}\eta
\beta^{\frac{d}{\eta}}}\left(  1-e^{-b_{n}n}\right) \\
& \sim\frac{8\sqrt{2\pi}\left(  \cosh2-3\right)  ^{2}\xi_{n}^{\frac{d}{\eta
}-1}}{e^{4}\eta\beta^{\frac{d}{\eta}}}n^{\frac{d}{\eta}-\frac{1}{2}}\left(
1+O\left(  n^{-\frac{1}{2}}\right)  \right)
\end{align*}
for some $\xi_{n}\in\left(  0,1\right)  $. Thus, for $d\geq\eta$, $\left\vert
\epsilon_{-}\left(  n\right)  \right\vert =O\left(  n^{\frac{d}{\eta}-\frac
{1}{2}}\right)  $.

For $d<\eta$, we have%
\begin{align*}
\left\vert \epsilon_{-}\left(  n\right)  \right\vert  & \leq\frac
{a_{n}e^{-b_{n}n}}{\eta\beta^{\frac{d}{\eta}}}\left(  \int_{0}^{1}x^{\frac
{d}{\eta}-1}e^{b_{n}x}\mathrm{d}x+\int_{1}^{n}x^{\frac{d}{\eta}-1}e^{b_{n}%
x}\mathrm{d}x\right) \\
& \leq\frac{a_{n}e^{-b_{n}\left(  n-1\right)  }}{d\beta^{\frac{d}{\eta}}%
}+\frac{a_{n}e^{-b_{n}n}}{\eta\beta^{\frac{d}{\eta}}}\int_{1}^{n}%
\underbrace{x^{\frac{d}{\eta}-1}e^{b_{n}x}}_{I_{n}\left(  x\right)
}\mathrm{d}x.
\end{align*}
The first term on the right hand side of this inequality goes to zero
exponentially with increasing $n$. Consequently, we focus on the second term.
Notice that $I_{n}\left(  x\right)  $ is convex, with a minimum at $x=\left(
1-\frac{d}{\eta}\right)  /b_{n}$. Moreover, $I_{n}\left(  1\right)
\rightarrow1$ as $n\rightarrow\infty$. Thus, the major contribution to the
integral $\int_{1}^{n}I_{n}$ occurs near the upper end of the integration
region, \emph{i.e.}, $x\approx n$. To make progress, let $y\left(  x\right)
=b_{n}x-\left(  1-\frac{d}{\eta}\right)  \log x$ such that $I_{n}\left(
x\right)  =e^{y\left(  x\right)  }$ and expand $y$ to first order about $x=n$.
This yields the accurate approximation%
\[
I_{n}\left(  x\right)  \approx e^{y\left(  n\right)  +y^{\prime}\left(
n\right)  \left(  x-n\right)  }.
\]
We can now integrate easily to obtain%
\[
\int_{1}^{n}I_{n}\left(  x\right)  \mathrm{d}x\approx\frac{e^{1-\frac{d}{\eta
}}n^{\frac{d}{\eta}}}{b_{n}n+\frac{d}{\eta}-1}\left(  e^{b_{n}n+\frac{d}{\eta
}-1}-e^{b_{n}-\frac{1-\frac{d}{\eta}}{n}}\right)
\]
and thus, to a good approximation, we can write%
\begin{align*}
\left\vert \epsilon_{-}\left(  n\right)  \right\vert  & \leq\frac
{a_{n}e^{-b_{n}\left(  n-1\right)  }}{d\beta^{\frac{d}{\eta}}}+\frac
{a_{n}e^{1-\frac{d}{\eta}}n^{\frac{d}{\eta}-1}}{\eta\beta^{\frac{d}{\eta}%
}b_{n}\left(  1+\frac{\frac{d}{\eta}-1}{b_{n}n}\right)  }\left(  e^{\frac
{d}{\eta}-1}-e^{-b_{n}\left(  n-1\right)  -\frac{1-\frac{d}{\eta}}{n}}\right)
\\
& \sim\frac{8\sqrt{2\pi}\left(  \cosh2-3\right)  ^{2}}{e^{4}\eta\beta
^{\frac{d}{\eta}}}n^{\frac{d}{\eta}-\frac{1}{2}}\left(  1+O\left(
n^{-\frac{1}{2}}\right)  \right)
\end{align*}
and so we have shown that $\epsilon\left(  n\right)  =O\left(  n^{\frac
{d}{\eta}-\frac{1}{2}}\right)  $.



\end{document}